\documentclass[superscriptaddress,prl,twocolumn,amssymb,nofootinbib,aps]{revtex4}
\usepackage{graphicx,bm,times}
\usepackage{tabularx} 
\usepackage{lipsum}
\usepackage{amsmath}
\usepackage{amssymb}
\usepackage{gensymb}
\usepackage{bbold}
\usepackage{float}
\usepackage{color,soul, mathrsfs}
 \usepackage{relsize}
  \newcounter{daggerfootnote}
\newcommand*{\daggerfootnote}[1]{%
    \setcounter{daggerfootnote}{\value{footnote}}%
    \renewcommand*{\thefootnote}{\fnsymbol{footnote}}%
    \footnote[2]{#1}%
    \setcounter{footnote}{\value{daggerfootnote}}%
    \renewcommand*{\thefootnote}{\arabic{footnote}}%
    }

\begin{document}

\title{
Twin-beam sub-shot-noise raster-scanning microscope with a hybrid detection scheme
}

\author{J.~Sabines-Chesterking}
\affiliation{Quantum Engineering Technology Labs, H. H. Wills Physics Laboratory and Department of Electrical \& Electronic Engineering, University of Bristol, BS8 1FD, UK.}

\affiliation{Joint Quantum Institute, National Institute of Standards and Technology and University of
Maryland, Gaithersburg, Maryland 20899, USA}
\author{A.~R.~McMillan}
\affiliation{Quantum Engineering Technology Labs, H. H. Wills Physics Laboratory and Department of Electrical \& Electronic Engineering, University of Bristol, BS8 1FD, UK.}
\author{P.~A.~Moreau}
\affiliation{Quantum Engineering Technology Labs, H. H. Wills Physics Laboratory and Department of Electrical \& Electronic Engineering, University of Bristol, BS8 1FD, UK.}
\affiliation{School of Physics and Astronomy, University of Glasgow, G12 8QQ, UK.}
\author{S.~K.~Joshi}
 \affiliation{Quantum Engineering Technology Labs, H. H. Wills Physics Laboratory and Department of Electrical \& Electronic Engineering, University of Bristol, BS8 1FD, UK.}
 \author{S.~Knauer}
  \affiliation{Quantum Engineering Technology Labs, H. H. Wills Physics Laboratory and Department of Electrical \& Electronic Engineering, University of Bristol, BS8 1FD, UK.}
 \affiliation{Centre for Quantum Computation \& Communication Technology,
School of Electrical Engineering \& Telecommunications,
University of New South Wales, Sydney, New South Wales 2052, Australia.}
  \author{E.~Johnston }
 \author{J.~G.~Rarity}
\author{J.~C.~F.~Matthews}
\affiliation{Quantum Engineering Technology Labs, H. H. Wills Physics Laboratory and Department of Electrical \& Electronic Engineering, University of Bristol, BS8 1FD, UK.}

\begin{abstract}
 \noindent 
By exploiting the quantised nature of light, we demonstrate a sub-shot-noise scanning optical transmittance microscope. Our microscope demonstrates, with micron scale resolution, a factor of improvement in precision of 1.76(9) in transmittance estimation gained per probe photon relative to an optimal classical version at the same optical power. This would allow us to observe photosensitive samples with nearly twice the precision, without sacrificing image resolution or increasing optical power to improve signal-to-noise ratio. Our setup uses correlated twin-beams produced by parametric down-conversion, and a hybrid detection scheme comprising photon-counting-based feed-forward and a highly efficient CCD camera.

\end{abstract}

\date{\today}

\maketitle


Imaging that harnesses the quantum properties of light promises transformative new capabilities, including imaging an object's interaction with one wavelength of light but by detecting at another distinct wavelength~\cite{lemos2014quantum}, imaging below the diffraction limit~\cite{giovannetti2009sub}, and reducing the uncertainty in the pointing of a laser beam~\cite{tr-sci-301-940}. Quantum states of light are also known to enable higher precision absorption imaging~\cite{br-nphot-4-227,samantaray2017realization,li2018enhanced} and phase imaging~\cite{on-natcomm-4-2426,PhysRevLett.112.103604} relative to ideal classical light sources, which are ultimately limited by optical shot-noise. Sub-shot-noise measurements are particularly important for enhanced imaging and measurement of light sensitive photo-reactive  biological samples, as higher precision measurements can provide better quality images for a given limited level of probe beam intensity~\cite{taylor2016quantum}.

The shot-noise of an ideal laser limits the minimum amplitude noise of an ideal classical light source, which is described by a Poisson distribution~\cite{WallsandMilburn}. Several reported 
approaches are able to surpass this limit, including the use of cooled and regularly pumped semiconductor lasers~\cite{PhysRevLett.58.1000, PhysRevLett.64.400}, generation of quadrature squeezed 
light~\cite{Andersen_2016}, and using entangled Fock states~\cite{Slussarenko:2017aa}. For the measurement of transmittance, the most precise sub-shot-noise approach per photon flux is the use of Fock states as a probe~\cite{Adesso09}. These states can be readily generated by exploiting the photon-number correlations from twin-beam parametric processes such as spontaneous parametric down conversion (SPDC)~\cite{hong1986experimental} or spontaneous four-wave mixing (SFWM)~\cite{levenson1985generation}.

For photon pair generation processes, the strong photon-number correlations between each of the twin-beams enable intensity fluctuations of the probe light to be characterized or suppressed~\cite{heidmann1987observation}, which in turn enables a higher signal-to-noise ratio of the mean photon number of the beam passing through the sample for parameter estimation. However, optical loss degrades the intensity correlation of the generated light, such that this technique can only provide an advantage beyond classical techniques when the total loss is reduced below specific thresholds~\cite{Jakeman86,br-nphot-4-227}. Previous demonstrations of quantum sensing and imaging that did not satisfy these thresholds required postselection of successful detection events to observe the physics of sub-shot-noise performance~\cite{whittaker2017absorption,li2018enhanced}, which underestimates the light exposure of the imaged sample.

Demonstrations of twin-beam sub-shot-noise single channel absorption estimation~\cite{moreau2017demonstrating} showed how commercially available high quantum efficiency 
cooled CCD cameras enable precision beyond the ideal classical scenario where Poisson-distributed light is detected with a perfect 100\% efficient detector. It has also been shown that fast optical gating, conditioned on single-photon detection events, can allow heralded photon counting twin-beam experiments to achieve sub-shot-noise absorption estimation without postselection~\cite{sabines2017sub}, even when the components of the system have overall efficiencies below the required thresholds that would otherwise require post-selection. Here we combine optical gating of single-photons and detection using a low-noise, high-efficiency CCD camera. 

Combining these schemes increases the available quantum advantage offered by the individual constituent detection schemes and enables the use of photon-coincidence-detection-based estimators with CCD camera intensity detection. The single-photon counting element allows for the removal
of virtually all classically contributed fluctuations in the light illuminating the sample~\cite{sabines2017sub}. This approach
contrasts with methods based purely on intensity correlations
where in practice the sample is illuminated with light comprising fluctuations that are partially of classical origin, due, for example, to losses in the reference photon channel.
In the presence of such classical fluctuations, the best known intensity correlation strategy to extract maximum information about the sample is to use estimators that exploit both the quantum and the classical contributions to the light intensity fluctuations~\cite{moreau2017demonstrating}. However, compared to photon counting, this strategy
is sub-optimal because firstly, the classical contribution will be fundamentally limited by the shot-noise, and secondly it causes the estimator to be sensitive to technical fluctuations (that can be super-Poissonian) in the source intensity.
In contrast, the new method reported here harnesses both the high efficiency of linear (non Geiger-mode) detectors and the classical noise suppression capability of optical gating with fast switching, conditional on single-photon detection events.

Many metrology applications seek to maximise measurement precision, which is statistically defined as the inverse of the variance of an experimentally estimated parameter. The variance is proportional to the uncertainty of experimental measurements used in the estimator. Here we focus on minimising the noise introduced into transmittance estimation caused by intensity fluctuations of an optical probe. 
 
Fock states can be prepared from correlated twin-beams of either SPDC or SFWM, by separating signal and idler light into two modes deterministically, most commonly via wavelength or polarization selection.
By detecting a photon-number state (such as a single-photon) in one of the beams, the presence of its companion is heralded in the other beam and thus then be used as an optical probe. Any loss in the output Fock state's path degrades the heralded state into a mixture. However, this mixture still has a smaller photon-number variance than that of Poisson-distributed light with the same average intensity--- it therefore can still be used to outperform classical light in transmittance estimation.

If the detector can resolve the time of arrival of single-photons, then single-photon coincidence rates can be used to estimate transmittance of a sample using the ratio between the coincidence counts and the single counts of the reference photons ($\langle N_\text{C}\rangle\langle/ N_\text{R} \rangle$). This parameter is also know as the heralding or Klyshko efficiency \cite{klyshko1980use}. To estimate the transmittance of a sample, $\eta_\text{S}$, it would only be necessary to obtain the ratio of the of Klyshko efficiency of the probe path with ($\eta_\text{{P'}}$) and without ($\eta_\text{{P}}$) the sample in place~\cite{Jakeman86}:
\begin{equation}
\hat{\eta}_\text{S}=\hat{\eta}_\text{P}/\hat{\eta}_\text{{P'}}=\frac{\langle N_\text{C}\rangle}{\langle N_\text{R} \rangle}/\frac{\langle N'_\text{C}\rangle}{\langle N'_\text{R} \rangle}.
\label{d}
\end{equation}
 Here $N_\text{C}$ is the number of coincidence counts between the reference and probe beam and $N_\text{R}$ is the number of single counts of the reference beam.
Performance 
is then compared to that of a direct transmittance measurement with 
shot-noise limited light
by estimating the ratio, $\Gamma$, 
of the precision obtained with each strategy. Sub-shot-noise (SSN) performance is indicated by $\Gamma>1$.
When the transmittance of the sample and the Klyshko efficiency of both the probe and reference channel are known, $\Gamma$ can be estimated using
\begin{equation}
\Gamma=\eta_\text{R}/({1-\eta_\text{S}\eta_\text{P}}).
\label{gamma}
\end{equation}
From Eq.~(\ref{gamma}), we can see that SSN performance can only be achieved when
$\eta_\text{P} + \eta_\text{R}>1$, where $\eta_\text{R}$ is the Klyshko efficiency of the reference arm \cite{Jakeman86}.

In order to use the estimator in Eq.~(\ref{d}), we use a single-photon avalanche detector (SPAD) to implement feed-forward with an optical delay and a fast optical switch so that probe photons are optically gated. This exposes the sample only when a herald photon was successfully detected, effectively reducing the losses of the reference channel and increasing the value of its Klyshko efficiency, $\eta_{\text{R}}$, closer to 1, making it easier to satisfy the aforementioned condition for obtaining an advantage, as demonstrated in~\cite{brida2012extremely, sabines2017sub}. We then use a CCD camera to detect the heralded probe photons. 
Since probe photons only reach the sample after being heralded, the number of photons that reach the CCD in the ideal case equals the number of photon coincidence events $\langle N_C\rangle$. Furthermore, using a modern CCD camera as a detector allows us to reach detection efficiencies that rival (albeit having more noise) cryogenically-cooled superconducting nanowire single-photon detectors~\cite{marsili2013detecting}, which have recently been used to achieve post-selection-free sub-shot-noise phase estimation with NOON states~\cite{Slussarenko:2017aa}. 
%
\begin{figure}
 \includegraphics[width=0.9\columnwidth]{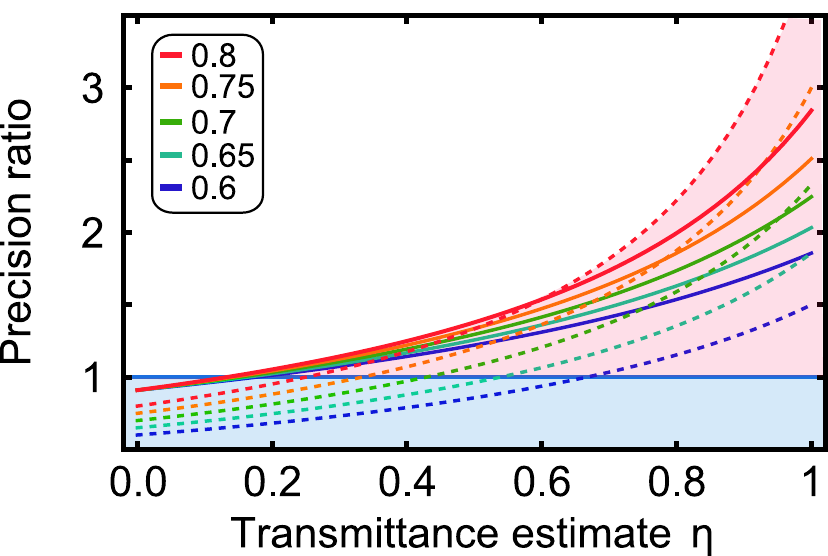}
 \centering
 \caption[Simulation of precision ratio using twin-beams and feed-forward.]{\textbf{Simulation of precision ratio using twin-beams and feed-forward.} The plot compares Poisson-limited light
 (horizontal blue line at ratio of 1) to transmittance estimation with twin-beam strategies. The key refers to the different values of overall source Klyshko efficiency, which are symmetric for direct twin-beam exposure (dotted lines) and asymmetric for feed-forward (solid lines) due to the 15$\%$ loss in the switch and 10$\%$ leakage of unheralded photons.} 
 \label{fd}
\end{figure}
\begin{figure*}
\centering
\includegraphics[width=0.9 \textwidth]{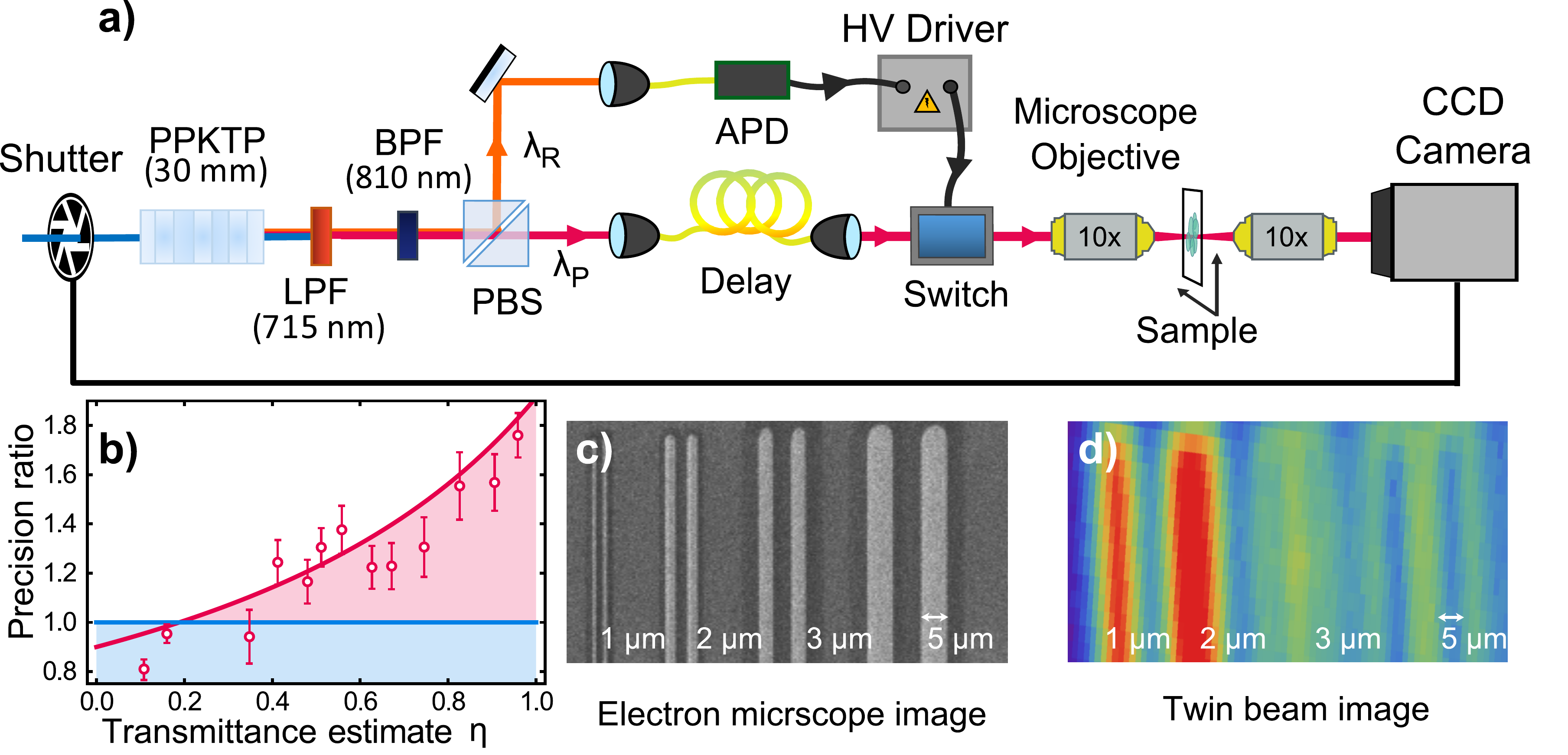}
\caption{\textbf{Experimental setup and its characterisation.} a) Photon-pairs are generated by SPDC in a PPKTP crystal, feed-forward of the idler photons is implemented using a polarisation independent switch, as described in Ref.~\cite{sabines2017sub}. Heralded photons are focused first through the sample and subsequently refocused to a point on a CCD camera for detection. The sample is raster scanned in two dimensions, allowing an image to be reconstructed. A mechanical shutter before the crystal is triggered by activation of the camera sensor, initializing the experiment and allowing the pump beam into the crystal. The shutter prevents undesired exposure of the sample during the camera's dead time. The acquisition time of the SPAD is synchronized to the CCD using a software interface that communicates with a time tagging electronics system which only records counts during the acquisition time of the camera, minimising uncorrelated reference channel noise. b) Precision ratio of transmittance between our setup and a theoretically ideal Poisson-limited light source
 measured with the same detection efficiency, (horizontal blue line at a ratio of 1, which corresponds to the shot-noise limit). Data points are the ratio between the variance of our transmittance estimate and the calculated variance of the same measurement for an ideal coherent state, at different levels of sample transmittance (red circles). In each case, the data was normalized relative to the number of input photons incident on the sample. Points above the blue line exhibit sub-shot-noise performance. Each point corresponds to 13 series of 40 measurements with an integration time of 1 s at a rate of 40000 counts per second. Error bars correspond to the experimental standard deviation of the mean from the different data series. c) Electron beam microscope image of the platinum deposited target used for resolution characterisation. d) Results of imaging the resolution target using correlated photon pairs. Two distinct strips are apparent in the optical image for 3~$\mu$m spacing, but for smaller spacings the gap between parallel strips is not discernible. From this, we estimate that the spatial resolution of the SSN microscope to be 3~$\mu$m. }
 \label{setupimaging}
\end{figure*}

When considering realistic, imperfect components, any leakage of unheralded photons through the switch due to a poor extinction ratio will reduce the Klyshko efficiency of the reference channel $\eta_{\text{R}}<1$, while extra loss introduced in optical delays and optical switches
will reduce $\eta_{\text{P}}$ . Therefore, when introducing an optical switch, there is a trade-off between increasing the Klyshko efficiency of the reference channel and introducing loss on the probe channel. Fig.~\ref{fd} shows the precision ratio that can be obtained by implementing feed-forward on twin-beam sources with different levels of Klyshko efficiencies. For illustration, we have assumed that the loss on the probe beam, due to the feed-forward optics, is the same as that measured in our setup (15$\%$) and has the same level of leakage of unheralded photons (10$\%$). In Fig.~\ref{fd} we observe, that feed-forward allows SSN performance to be achieved at lower transmittance than without feed-forward. It is also apparent that for Klyshko efficiencies below 70$\%$, feed-forward enables a higher precision ratio than directly using the correlated twin-beam source without optical gating.

The SSN microscope setup is shown in Fig.~\ref{setupimaging}, which is adapted from the 
photon-pair source and optical switching circuit used in~\cite{sabines2017sub,moreau2017demonstrating}. Correlated photon pairs are generated by type II collinear SPDC in a 30 mm periodically poled potassium titanyl phosphate crystal (PPKTP) pumped with a continuous wave (CW) 404 nm laser (Toptica, TOPMode 405~HP\protect\daggerfootnote{Disclaimer: Commercial equipment, instruments or materials are identified in this report to foster understanding. Such identification does not imply recommendation or endorsement by the National Institute of Standards and Technology, nor does it imply that the materials or equipment are necessarily the best available for the purpose.\label{ftn:X}}). The wavelengths of the down-converted photons are tuned by adjusting the temperature of the crystal such that the correlated beams are emitted at 818 nm (probe photons) and 798 nm (reference photons). Generated photon pairs are filtered using a long pass and a band pass filter and then deterministically separated using a polarizing beam splitter (PBS). After being separated, the photons are each coupled into separate single-mode fibers. The probe photon is then detected with a SPAD, heralding the presence of its companion and triggering an optical switch in the reference photon channel. 

The optical switch is formed by a free-space Pockels cell modulator (Thorlabs, EO-AM-NR-C1\textsuperscript{\ref{ftn:X}}) inside a Sagnac loop as described in \cite{sabines2017sub}. When the optical switch has been triggered by detection of a probe photon, the corresponding reference photon is transmitted through the switch, focused onto the sample and then recollimated after the sample using a pair of 10x microscope objectives. Once the reference beam has traversed the sample and has been collimated it is then refocused onto the sensor of a CCD camera (Andor, iDus 416\textsuperscript{\ref{ftn:X}}) for detection. The sample is raster scanned in the plane perpendicular to the path of the beam using a pair of electronically driven linear stages. As the intensity of the light passing through the sample is measured, we reconstruct an image of the sample transmittance on a point by point basis. 

\begin{figure*}[t]
\centering
\includegraphics[width=0.9 \textwidth]{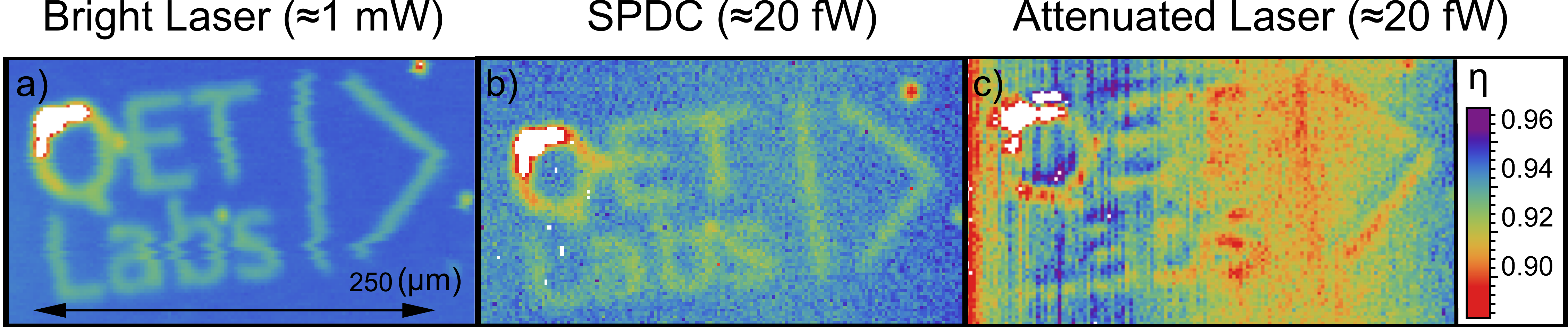}
 \caption[Experimental Results]{
\textbf{Experimental Results.} a) Reference image of the sample, acquired with differential imaging and 1~mW
of laser light passing through the sample. b) Noise-reduced image using correlated photon pairs at a rate of 40000 per second
passing through the sample, which is equivalent to 20 fW of light passing through the sample. 
c) A differential image taken with a laser attenuated to the same rate of photons as used in b). Each image is 150 by 75 pixels. The step size of each pixel is 2 $\mu$m and the integration time per pixel for a) is 0.1 s and for b) and c) is 1 s. White regions correspond to low transmittance values that are below the scale.}
\label{qimage}
\end{figure*}

The sample we imaged is a 3 mm thick AR coated N-BK7 window, in which features were engraved using ion-beam-milling. By selectively thinning the anti-reflection coating of the window a low contrast transmittance figure was created. The ion-beam-milled figures have high transmittance ($\approx$95$\%$) and low contrast ($\approx$2$\%$) which make them ideal to show the practicality of the twin-beam imaging scheme. In other sections of the sample, platinum deposition was used to create higher contrast markers for locating the low contrast sample and to provide targets with which to benchmark resolution. A high power reference image from the same imaging setup (Fig.~\ref{qimage}~a), was taken by raster scanning a bright laser set at 1 mW of power (at the target), and measuring the transmitted power with a photodiode power meter.



We characterized the resolution of our imaging system using a resolution target consisting of a series of parallel lines with widths ranging from 5 $\mu$m to 1 $\mu$m (Fig.~\ref{setupimaging}~b). Using the correlated photon pairs to image, we resolve features with widths of down to 3~$\mu$m, (Fig.~\ref{setupimaging}~c).
%

The 
leakage of unheralded photons through the optical switch was characterized
by measuring the Klyshko efficiency $\eta_\text{R}$ of the reference beam using SPADs in both channels. We obtained 
$\eta_\text{R}=N_\text{C}/N_\text{P}=90(3)\%$, which corresponds to the proportion of photons exposed to the sample that have been successfully heralded, and
therefore the percentage of the photons detected on the camera can be treated as coincidences for our transmittance estimator. 

We next characterized the system's performance for precision estimation
using a variable neutral-density
filter to act as an absorptive sample and estimated the transmittance for different values of attenuation. Using this procedure, we 
compared our system's precision in estimating transmittance $1/\Delta^2\eta_{\text{Exp}}$,  (normalised by the mean intensity of the probe beam $\langle N_{\text{P}_{\text{In}}}\rangle$) to that of an ideal
shot-noise limited
scheme, $1/\Delta^2\eta_{\text{Coh}}$, measured with a detector which has the same efficiency as the one used in this experiment (90$\%$).   
\begin{equation}
\Gamma={\Delta^2\eta_{\text{Coh}}}/{(\Delta^2\eta_{\text{Exp}}\,\,\langle N_{P_{\text{In}}}\rangle)}.
\end{equation}
Here the number of probe photons is 
estimated from the
mean number of detected probe photons, $\langle N_{\text{P}_{\text{Det}}}\rangle$, corrected by the CCDs dark counts, $\langle N_{\text{DC}}\rangle$, the transmittance of the optics after the sample, $\eta_{\text{Opt}}$, and the transmittance of the sample, $\eta$, 
\begin{equation}
\langle N_{P_{\text{IN}}}\rangle=({\langle N_{P_{\text{Det}}}\rangle}-\langle N_{\text{DC}}\rangle)/\left(\eta_{\text{Opt}}\eta\right).
\end{equation}
The results of the system performance characterization are shown in Fig.~\ref{setupimaging}~b.

The maximum precision ratio obtained while characterizing our setup without a sample when compared with a direct measurement classical scheme using the same detector efficiency (90$\%$) was $\Gamma=$1.76(9), the reported uncertainty is given by the experimental standard deviation of the mean from the different series of data that were acquired. The precision ratio rises to $\Gamma_{\text{Dif}}=$2.52(9) when compared to an ideal classical differential scheme. The system achieves sub-shot-noise performance for a transmittance higher than 0.4. Notably this system shows an absolute precision ratio of  $\Gamma_{\text{Abs}}=$ 1.58(9) over an ideal 100$\%$ detection efficiency classical experiment.


In Fig.~\ref{qimage} we compare images obtained using correlated photon pairs in our microscopy setup to those obtained using classical illumination. For the classical images (Fig.~\ref{qimage}~c), we used a laser beam, attenuated to the same level of intensity as the probe photons input to the microscope from the SPDC source, together with a 50$\%$ reflective beamsplitter inserted after the optical switch and used the reflection to monitor power fluctuations on an additional SPAD, implementing a low power differential measurement. Fig.~\ref{qimage}~a shows a reference image taken with the same method, but with the laser turned to 1~mW, and using two power meters as the detectors.

\begin{figure}[tp!]
 \centering
 \includegraphics[width=0.85\columnwidth]{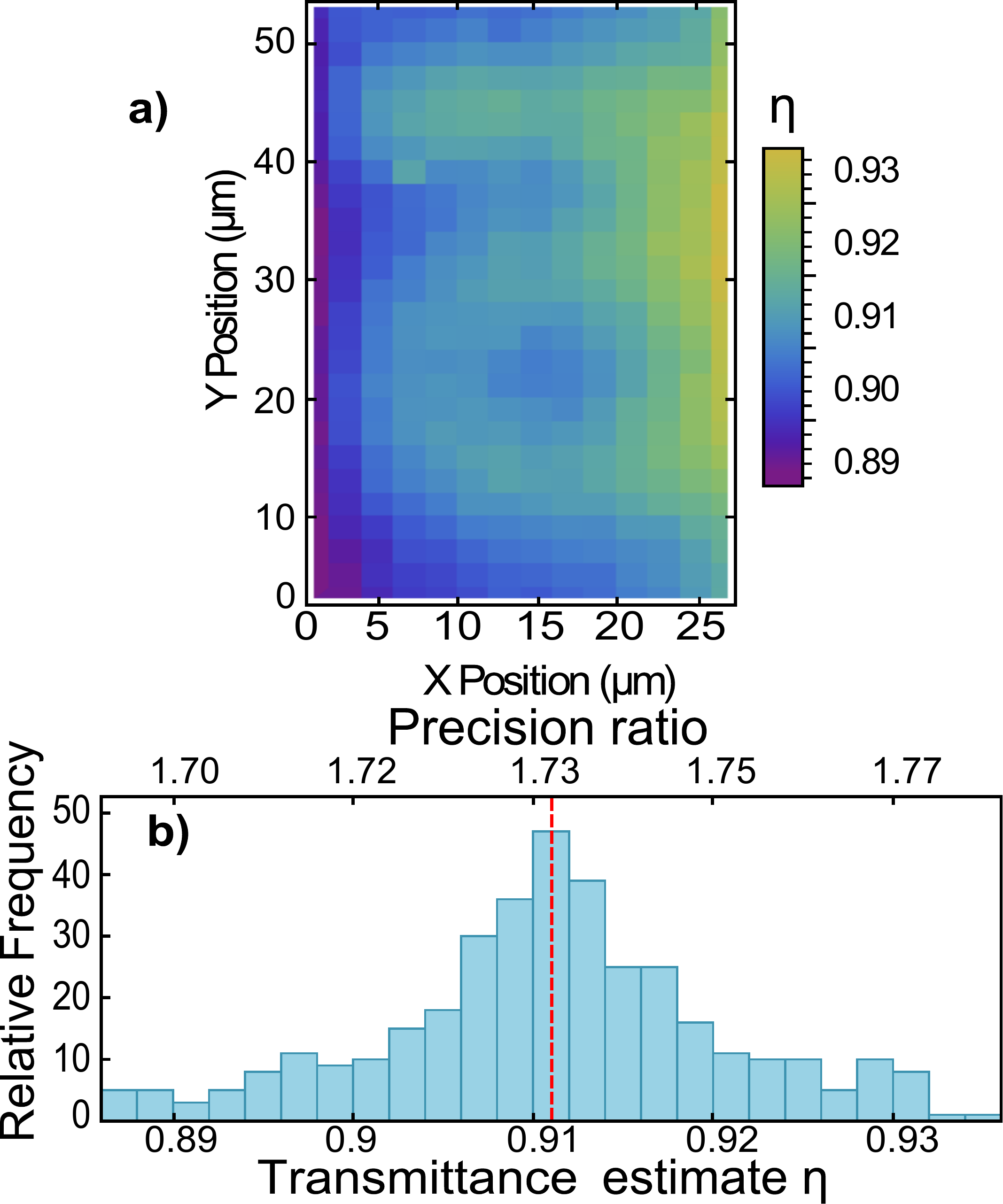}
 \caption[Multiple scan image]
 {
\textbf{Multiple scanned image analysis} 
a) A 14x26 pixels subsection of the object imaged in Fig.~\ref{qimage} in the vicinity of letter ``a". The step size between pixels is 2 $\mu$m. Each pixel corresponds to the mean of 80 recorded measurements with an integration time of 1 s. The gradient between the left and right side of the image is due to slow drift in the source brightness. b) Histogram of the transmittance and expected precision ratio.  The red vertical dashed lines corresponds to the mean transmittance $\eta=$0.911 and the bin width is 20.}
 \label{a}
\end{figure}

The time taken to acquire the low light intensity images in Fig.~\ref{qimage}~(b,c) ($\approx$ 1 s per pixel) in our current setup, precluded obtaining a statistically valid number of copies of one image to allow estimation of the variance between them. Instead we imaged eighty copies of a subset of the sample, at the same resolution, and computed the variance across them in a pixel-by-pixel analysis---the data for these images is shown in Fig.~\ref{a}. 
When the variance across this sample is compared to the variance that would be obtained with a Poisson distributed illumination source we find a mean precision ratio of $\Gamma=$1.54(26), which lies close to the expected value of $\Gamma=$1.73 calculated from the calibration curve in 
Fig.~\ref{setupimaging}~d.
When we compare to a Poisson-limited source measured with the a 100$\%$ efficient detector, we find  $\Gamma_{\text{Abs}}=$ 1.39(26).
As the transmittance is not homogeneous across the sample, the expected advantage in precision varies across the image. Fig.~\ref{a}
shows a histogram of the transmittance of the eighty images used to translate transmittance into the expected precision ratio using 
Fig.~\ref{setupimaging}~d, indicated by the top x-axis.

We have implemented a 3~$\mu$m resolution scanning transmittance microscope that operates with precision that is up to $76\%$ better than the shot-noise limit obtained using a classical (shot noise limited) light source in an equivalent apparatus. We implemented this microscope with a 
SPDC photon pair source together with single-photon feed-forward in a hybrid CCD-SPAD detection scheme, which requires no active stabilisation or locking of cavity optics. The hybrid detection method we use allows data gathered with a CCD---that by itself is incapable of acquiring photon arrival information---to be used in a coincidence-based estimator of transmittance, at the efficiency achievable in commercially available cooled-CCD sensors. Similar performance might be achieved using a high efficiency superconducting photon-counting detector~\cite{marsili2013detecting}, but this would come at the financial, space and power costs of current implementations of superconducting technology. 
%
%
Our analysis accounts for all photons passing through the sample, for both our measured photon pair experiment and the comparison to an idealised classical experiment. 

In comparison with the wide field microscope presented in~\cite{samantaray2017realization}, our imaging system is slow due to the need to raster scan the sample. However, because we emit the probe beam from a single-mode fibre, it is straightforward for the microscope to achieve a high spatial resolution (down to 3 $\mu$m) close to the diffraction limit, whilst simultaneously maintaining a high quantum advantage. The resolution could be improved by increasing the numerical aperture  of the confocal imaging lenses, in principle to about 0.4$\mu$m (half a wavelength) but would come at the cost of increased alignment sensitivity due to reduced depth of focus.

We believe that our scheme could be readily applied to scenarios where it is desired to image a sample with high resolution at the single-photon intensity level. For example, it could be used to obtain a precise spatial characterization of a single-photon detector like the CCD sensor, or equivalently in single-photon electrophysiology scenarios~\cite{frogexperiment}, where high precision targeted delivery of single-photons, and high resolution spatial information of responses may be required. However, a common critique of twin-beam noise reduction schemes operating within the photon counting regime is that their inherent low intensity is well below any damage threshold of realistic samples---we speculate
our approach could be extended to higher intensities 
by combining optical delay and fast switching with the technique reported in~\cite{iskhakov2016heralded}, to post-select low noise high intensity beams, increasing
the intensity on the sample, whilst maintaining sub-shot-noise performance.

\section{ACKNOWLEDGMENTS}
We thank Alan Migdall for his comments and constructive criticism of the manuscript. This work was supported by EPSRC programme grant EP/L024020/1, EPSRC UK Quantum Technology Hub in Quantum Enhanced Imaging (EP/M01326X/1) and the Centre for Nanoscience and Quantum Information (NSQI). JGR acknowledges support from an EPSRC Quantum Technology Fellowship (EP/M024458/1).
JCFM acknowledges support from an EPSRC Quantum Technology Fellowship (EP/M024385/1) and an ERC starting grant ERC-2018-STG 803665.


\begin{thebibliography}{27}
\expandafter\ifx\csname natexlab\endcsname\relax\def\natexlab#1{#1}\fi
\expandafter\ifx\csname bibnamefont\endcsname\relax
  \def\bibnamefont#1{#1}\fi
\expandafter\ifx\csname bibfnamefont\endcsname\relax
  \def\bibfnamefont#1{#1}\fi
\expandafter\ifx\csname citenamefont\endcsname\relax
  \def\citenamefont#1{#1}\fi
\expandafter\ifx\csname url\endcsname\relax
  \def\url#1{\texttt{#1}}\fi
\expandafter\ifx\csname urlprefix\endcsname\relax\def\urlprefix{URL }\fi
\providecommand{\bibinfo}[2]{#2}
\providecommand{\eprint}[2][]{\url{#2}}

\bibitem[{\citenamefont{Lemos et~al.}(2014)\citenamefont{Lemos, Borish, Cole,
  Ramelow, Lapkiewicz, and Zeilinger}}]{lemos2014quantum}
\bibinfo{author}{\bibfnamefont{G.~B.} \bibnamefont{Lemos}},
  \bibinfo{author}{\bibfnamefont{V.}~\bibnamefont{Borish}},
  \bibinfo{author}{\bibfnamefont{G.~D.} \bibnamefont{Cole}},
  \bibinfo{author}{\bibfnamefont{S.}~\bibnamefont{Ramelow}},
  \bibinfo{author}{\bibfnamefont{R.}~\bibnamefont{Lapkiewicz}},
  \bibnamefont{and}
  \bibinfo{author}{\bibfnamefont{A.}~\bibnamefont{Zeilinger}},
  \bibinfo{journal}{Nature} \textbf{\bibinfo{volume}{512}},
  \bibinfo{pages}{409} (\bibinfo{year}{2014}).

\bibitem[{\citenamefont{Giovannetti et~al.}(2009)\citenamefont{Giovannetti,
  Lloyd, Maccone, and Shapiro}}]{giovannetti2009sub}
\bibinfo{author}{\bibfnamefont{V.}~\bibnamefont{Giovannetti}},
  \bibinfo{author}{\bibfnamefont{S.}~\bibnamefont{Lloyd}},
  \bibinfo{author}{\bibfnamefont{L.}~\bibnamefont{Maccone}}, \bibnamefont{and}
  \bibinfo{author}{\bibfnamefont{J.~H.} \bibnamefont{Shapiro}},
  \bibinfo{journal}{Physical Review A} \textbf{\bibinfo{volume}{79}},
  \bibinfo{pages}{013827} (\bibinfo{year}{2009}).

\bibitem[{\citenamefont{Treps et~al.}(2003)\citenamefont{Treps, Grosse, Bowen,
  Fabre, Bachor, and Lam}}]{tr-sci-301-940}
\bibinfo{author}{\bibfnamefont{N.}~\bibnamefont{Treps}},
  \bibinfo{author}{\bibfnamefont{N.}~\bibnamefont{Grosse}},
  \bibinfo{author}{\bibfnamefont{W.~P.} \bibnamefont{Bowen}},
  \bibinfo{author}{\bibfnamefont{C.}~\bibnamefont{Fabre}},
  \bibinfo{author}{\bibfnamefont{H.-A.} \bibnamefont{Bachor}},
  \bibnamefont{and} \bibinfo{author}{\bibfnamefont{P.~K.} \bibnamefont{Lam}},
  \bibinfo{journal}{Science} \textbf{\bibinfo{volume}{301}},
  \bibinfo{pages}{940} (\bibinfo{year}{2003}).

\bibitem[{\citenamefont{Brida et~al.}(2010)\citenamefont{Brida, Genovese, and
  Rou~Berchera}}]{br-nphot-4-227}
\bibinfo{author}{\bibfnamefont{G.}~\bibnamefont{Brida}},
  \bibinfo{author}{\bibfnamefont{M.}~\bibnamefont{Genovese}}, \bibnamefont{and}
  \bibinfo{author}{\bibfnamefont{I.}~\bibnamefont{Rou~Berchera}},
  \bibinfo{journal}{Nature Photon.} \textbf{\bibinfo{volume}{4}},
  \bibinfo{pages}{227} (\bibinfo{year}{2010}).

\bibitem[{\citenamefont{Samantaray et~al.}(2017)\citenamefont{Samantaray,
  Ruo-Berchera, Meda, and Genovese}}]{samantaray2017realization}
\bibinfo{author}{\bibfnamefont{N.}~\bibnamefont{Samantaray}},
  \bibinfo{author}{\bibfnamefont{I.}~\bibnamefont{Ruo-Berchera}},
  \bibinfo{author}{\bibfnamefont{A.}~\bibnamefont{Meda}}, \bibnamefont{and}
  \bibinfo{author}{\bibfnamefont{M.}~\bibnamefont{Genovese}},
  \bibinfo{journal}{Light: Science \& Applications}
  \textbf{\bibinfo{volume}{6}}, \bibinfo{pages}{e17005} (\bibinfo{year}{2017}).

\bibitem[{\citenamefont{Li et~al.}(2018)\citenamefont{Li, Zou, Liu, Guo, Guo,
  and Ren}}]{li2018enhanced}
\bibinfo{author}{\bibfnamefont{M.}~\bibnamefont{Li}},
  \bibinfo{author}{\bibfnamefont{C.-L.} \bibnamefont{Zou}},
  \bibinfo{author}{\bibfnamefont{D.}~\bibnamefont{Liu}},
  \bibinfo{author}{\bibfnamefont{G.-P.} \bibnamefont{Guo}},
  \bibinfo{author}{\bibfnamefont{G.-C.} \bibnamefont{Guo}}, \bibnamefont{and}
  \bibinfo{author}{\bibfnamefont{X.-F.} \bibnamefont{Ren}},
  \bibinfo{journal}{Physical Review A} \textbf{\bibinfo{volume}{98}},
  \bibinfo{pages}{012121} (\bibinfo{year}{2018}).

\bibitem[{\citenamefont{Ono et~al.}(2013)\citenamefont{Ono, Okamoto, and
  Takeuchi}}]{on-natcomm-4-2426}
\bibinfo{author}{\bibfnamefont{T.}~\bibnamefont{Ono}},
  \bibinfo{author}{\bibfnamefont{R.}~\bibnamefont{Okamoto}}, \bibnamefont{and}
  \bibinfo{author}{\bibfnamefont{S.}~\bibnamefont{Takeuchi}},
  \bibinfo{journal}{Nature Communications} \textbf{\bibinfo{volume}{4}}
  (\bibinfo{year}{2013}).

\bibitem[{\citenamefont{Israel et~al.}(2014)\citenamefont{Israel, Rosen, and
  Silberberg}}]{PhysRevLett.112.103604}
\bibinfo{author}{\bibfnamefont{Y.}~\bibnamefont{Israel}},
  \bibinfo{author}{\bibfnamefont{S.}~\bibnamefont{Rosen}}, \bibnamefont{and}
  \bibinfo{author}{\bibfnamefont{Y.}~\bibnamefont{Silberberg}},
  \bibinfo{journal}{Phys. Rev. Lett.} \textbf{\bibinfo{volume}{112}},
  \bibinfo{pages}{103604} (\bibinfo{year}{2014}).

\bibitem[{\citenamefont{Taylor and Bowen}(2016)}]{taylor2016quantum}
\bibinfo{author}{\bibfnamefont{M.~A.} \bibnamefont{Taylor}} \bibnamefont{and}
  \bibinfo{author}{\bibfnamefont{W.~P.} \bibnamefont{Bowen}},
  \bibinfo{journal}{Physics Reports} \textbf{\bibinfo{volume}{615}},
  \bibinfo{pages}{1} (\bibinfo{year}{2016}).

\bibitem[{\citenamefont{Walls and Milburn}(2008)}]{WallsandMilburn}
\bibinfo{author}{\bibfnamefont{D.~F.} \bibnamefont{Walls}} \bibnamefont{and}
  \bibinfo{author}{\bibfnamefont{G.~J.} \bibnamefont{Milburn}},
  \emph{\bibinfo{title}{Quanutm Optics, 2nd Edition}}
  (\bibinfo{publisher}{Springer}, \bibinfo{year}{2008}).

\bibitem[{\citenamefont{Machida et~al.}(1987)\citenamefont{Machida, Yamamoto,
  and Itaya}}]{PhysRevLett.58.1000}
\bibinfo{author}{\bibfnamefont{S.}~\bibnamefont{Machida}},
  \bibinfo{author}{\bibfnamefont{Y.}~\bibnamefont{Yamamoto}}, \bibnamefont{and}
  \bibinfo{author}{\bibfnamefont{Y.}~\bibnamefont{Itaya}},
  \bibinfo{journal}{Phys. Rev. Lett.} \textbf{\bibinfo{volume}{58}},
  \bibinfo{pages}{1000} (\bibinfo{year}{1987}).

\bibitem[{\citenamefont{Richardson and Shelby}(1990)}]{PhysRevLett.64.400}
\bibinfo{author}{\bibfnamefont{W.~H.} \bibnamefont{Richardson}}
  \bibnamefont{and} \bibinfo{author}{\bibfnamefont{R.~M.}
  \bibnamefont{Shelby}}, \bibinfo{journal}{Phys. Rev. Lett.}
  \textbf{\bibinfo{volume}{64}}, \bibinfo{pages}{400} (\bibinfo{year}{1990}).

\bibitem[{\citenamefont{Andersen et~al.}(2016)\citenamefont{Andersen, Gehring,
  Marquardt, and Leuchs}}]{Andersen_2016}
\bibinfo{author}{\bibfnamefont{U.~L.} \bibnamefont{Andersen}},
  \bibinfo{author}{\bibfnamefont{T.}~\bibnamefont{Gehring}},
  \bibinfo{author}{\bibfnamefont{C.}~\bibnamefont{Marquardt}},
  \bibnamefont{and} \bibinfo{author}{\bibfnamefont{G.}~\bibnamefont{Leuchs}},
  \bibinfo{journal}{Physica Scripta} \textbf{\bibinfo{volume}{91}},
  \bibinfo{pages}{053001} (\bibinfo{year}{2016}).

\bibitem[{\citenamefont{Slussarenko et~al.}(2017)\citenamefont{Slussarenko,
  Weston, Chrzanowski, Shalm, Verma, Nam, and Pryde}}]{Slussarenko:2017aa}
\bibinfo{author}{\bibfnamefont{S.}~\bibnamefont{Slussarenko}},
  \bibinfo{author}{\bibfnamefont{M.~M.} \bibnamefont{Weston}},
  \bibinfo{author}{\bibfnamefont{H.~M.} \bibnamefont{Chrzanowski}},
  \bibinfo{author}{\bibfnamefont{L.~K.} \bibnamefont{Shalm}},
  \bibinfo{author}{\bibfnamefont{V.~B.} \bibnamefont{Verma}},
  \bibinfo{author}{\bibfnamefont{S.~W.} \bibnamefont{Nam}}, \bibnamefont{and}
  \bibinfo{author}{\bibfnamefont{G.~J.} \bibnamefont{Pryde}},
  \bibinfo{journal}{Nature Photonics} \textbf{\bibinfo{volume}{11}},
  \bibinfo{pages}{700} (\bibinfo{year}{2017}).

\bibitem[{\citenamefont{Adesso et~al.}(2009)\citenamefont{Adesso, Dell'Anno,
  De~Siena, Illuminati, and Souza}}]{Adesso09}
\bibinfo{author}{\bibfnamefont{G.}~\bibnamefont{Adesso}},
  \bibinfo{author}{\bibfnamefont{F.}~\bibnamefont{Dell'Anno}},
  \bibinfo{author}{\bibfnamefont{S.}~\bibnamefont{De~Siena}},
  \bibinfo{author}{\bibfnamefont{F.}~\bibnamefont{Illuminati}},
  \bibnamefont{and} \bibinfo{author}{\bibfnamefont{L.~A.~M.}
  \bibnamefont{Souza}}, \bibinfo{journal}{Phys. Rev. A}
  \textbf{\bibinfo{volume}{79}}, \bibinfo{pages}{040305R}
  (\bibinfo{year}{2009}).

\bibitem[{\citenamefont{Hong and Mandel}(1986)}]{hong1986experimental}
\bibinfo{author}{\bibfnamefont{C.}~\bibnamefont{Hong}} \bibnamefont{and}
  \bibinfo{author}{\bibfnamefont{L.}~\bibnamefont{Mandel}},
  \bibinfo{journal}{Physical Review Letters} \textbf{\bibinfo{volume}{56}},
  \bibinfo{pages}{58} (\bibinfo{year}{1986}).

\bibitem[{\citenamefont{Levenson et~al.}(1985)\citenamefont{Levenson, Shelby,
  Aspect, Reid, and Walls}}]{levenson1985generation}
\bibinfo{author}{\bibfnamefont{M.}~\bibnamefont{Levenson}},
  \bibinfo{author}{\bibfnamefont{R.}~\bibnamefont{Shelby}},
  \bibinfo{author}{\bibfnamefont{A.}~\bibnamefont{Aspect}},
  \bibinfo{author}{\bibfnamefont{M.}~\bibnamefont{Reid}}, \bibnamefont{and}
  \bibinfo{author}{\bibfnamefont{D.}~\bibnamefont{Walls}},
  \bibinfo{journal}{Physical Review A} \textbf{\bibinfo{volume}{32}},
  \bibinfo{pages}{1550} (\bibinfo{year}{1985}).

\bibitem[{\citenamefont{Heidmann et~al.}(1987)\citenamefont{Heidmann, Horowicz,
  Reynaud, Giacobino, Fabre, and Camy}}]{heidmann1987observation}
\bibinfo{author}{\bibfnamefont{A.}~\bibnamefont{Heidmann}},
  \bibinfo{author}{\bibfnamefont{R.}~\bibnamefont{Horowicz}},
  \bibinfo{author}{\bibfnamefont{S.}~\bibnamefont{Reynaud}},
  \bibinfo{author}{\bibfnamefont{E.}~\bibnamefont{Giacobino}},
  \bibinfo{author}{\bibfnamefont{C.}~\bibnamefont{Fabre}}, \bibnamefont{and}
  \bibinfo{author}{\bibfnamefont{G.}~\bibnamefont{Camy}},
  \bibinfo{journal}{Physical review letters} \textbf{\bibinfo{volume}{59}},
  \bibinfo{pages}{2555} (\bibinfo{year}{1987}).

\bibitem[{\citenamefont{Jakeman and Rarity}(1986)}]{Jakeman86}
\bibinfo{author}{\bibfnamefont{E.}~\bibnamefont{Jakeman}} \bibnamefont{and}
  \bibinfo{author}{\bibfnamefont{J.~G.} \bibnamefont{Rarity}},
  \bibinfo{journal}{Opt. Comm.} \textbf{\bibinfo{volume}{59}},
  \bibinfo{pages}{219} (\bibinfo{year}{1986}).

\bibitem[{\citenamefont{Whittaker et~al.}(2017)\citenamefont{Whittaker, Erven,
  Neville, Berry, O'Brien, Cable, and Matthews}}]{whittaker2017absorption}
\bibinfo{author}{\bibfnamefont{R.}~\bibnamefont{Whittaker}},
  \bibinfo{author}{\bibfnamefont{C.}~\bibnamefont{Erven}},
  \bibinfo{author}{\bibfnamefont{A.}~\bibnamefont{Neville}},
  \bibinfo{author}{\bibfnamefont{M.}~\bibnamefont{Berry}},
  \bibinfo{author}{\bibfnamefont{J.}~\bibnamefont{O'Brien}},
  \bibinfo{author}{\bibfnamefont{H.}~\bibnamefont{Cable}}, \bibnamefont{and}
  \bibinfo{author}{\bibfnamefont{J.}~\bibnamefont{Matthews}},
  \bibinfo{journal}{New Journal of Physics} \textbf{\bibinfo{volume}{19}},
  \bibinfo{pages}{023013} (\bibinfo{year}{2017}).

\bibitem[{\citenamefont{Moreau et~al.}(2017)\citenamefont{Moreau,
  Sabines-Chesterking, Whittaker, Joshi, Birchall, McMillan, Rarity, and
  Matthews}}]{moreau2017demonstrating}
\bibinfo{author}{\bibfnamefont{P.-A.} \bibnamefont{Moreau}},
  \bibinfo{author}{\bibfnamefont{J.}~\bibnamefont{Sabines-Chesterking}},
  \bibinfo{author}{\bibfnamefont{R.}~\bibnamefont{Whittaker}},
  \bibinfo{author}{\bibfnamefont{S.~K.} \bibnamefont{Joshi}},
  \bibinfo{author}{\bibfnamefont{P.~M.} \bibnamefont{Birchall}},
  \bibinfo{author}{\bibfnamefont{A.}~\bibnamefont{McMillan}},
  \bibinfo{author}{\bibfnamefont{J.~G.} \bibnamefont{Rarity}},
  \bibnamefont{and} \bibinfo{author}{\bibfnamefont{J.~C.}
  \bibnamefont{Matthews}}, \bibinfo{journal}{Scientific Reports}
  \textbf{\bibinfo{volume}{7}}, \bibinfo{pages}{6256} (\bibinfo{year}{2017}).

\bibitem[{\citenamefont{Sabines-Chesterking
  et~al.}(2017)\citenamefont{Sabines-Chesterking, Whittaker, Joshi, Birchall,
  Moreau, McMillan, Cable, O'Brien, Rarity, and Matthews}}]{sabines2017sub}
\bibinfo{author}{\bibfnamefont{J.}~\bibnamefont{Sabines-Chesterking}},
  \bibinfo{author}{\bibfnamefont{R.}~\bibnamefont{Whittaker}},
  \bibinfo{author}{\bibfnamefont{S.}~\bibnamefont{Joshi}},
  \bibinfo{author}{\bibfnamefont{P.}~\bibnamefont{Birchall}},
  \bibinfo{author}{\bibfnamefont{P.}~\bibnamefont{Moreau}},
  \bibinfo{author}{\bibfnamefont{A.}~\bibnamefont{McMillan}},
  \bibinfo{author}{\bibfnamefont{H.}~\bibnamefont{Cable}},
  \bibinfo{author}{\bibfnamefont{J.}~\bibnamefont{O'Brien}},
  \bibinfo{author}{\bibfnamefont{J.}~\bibnamefont{Rarity}}, \bibnamefont{and}
  \bibinfo{author}{\bibfnamefont{J.}~\bibnamefont{Matthews}},
  \bibinfo{journal}{Physical Review Applied} \textbf{\bibinfo{volume}{8}},
  \bibinfo{pages}{014016} (\bibinfo{year}{2017}).

\bibitem[{\citenamefont{Klyshko}(1980)}]{klyshko1980use}
\bibinfo{author}{\bibfnamefont{D.}~\bibnamefont{Klyshko}},
  \bibinfo{journal}{Soviet Journal of Quantum Electronics}
  \textbf{\bibinfo{volume}{10}}, \bibinfo{pages}{1112} (\bibinfo{year}{1980}).

\bibitem[{\citenamefont{Brida et~al.}(2012)\citenamefont{Brida, Degiovanni,
  Genovese, Piacentini, Traina, Della~Frera, Tosi, Bahgat~Shehata, Scarcella,
  Gulinatti et~al.}}]{brida2012extremely}
\bibinfo{author}{\bibfnamefont{G.}~\bibnamefont{Brida}},
  \bibinfo{author}{\bibfnamefont{I.}~\bibnamefont{Degiovanni}},
  \bibinfo{author}{\bibfnamefont{M.}~\bibnamefont{Genovese}},
  \bibinfo{author}{\bibfnamefont{F.}~\bibnamefont{Piacentini}},
  \bibinfo{author}{\bibfnamefont{P.}~\bibnamefont{Traina}},
  \bibinfo{author}{\bibfnamefont{A.}~\bibnamefont{Della~Frera}},
  \bibinfo{author}{\bibfnamefont{A.}~\bibnamefont{Tosi}},
  \bibinfo{author}{\bibfnamefont{A.}~\bibnamefont{Bahgat~Shehata}},
  \bibinfo{author}{\bibfnamefont{C.}~\bibnamefont{Scarcella}},
  \bibinfo{author}{\bibfnamefont{A.}~\bibnamefont{Gulinatti}},
  \bibnamefont{et~al.}, \bibinfo{journal}{Applied Physics Letters}
  \textbf{\bibinfo{volume}{101}}, \bibinfo{pages}{221112}
  (\bibinfo{year}{2012}).

\bibitem[{\citenamefont{Marsili et~al.}(2013)\citenamefont{Marsili, Verma,
  Stern, Harrington, Lita, Gerrits, Vayshenker, Baek, Shaw, Mirin
  et~al.}}]{marsili2013detecting}
\bibinfo{author}{\bibfnamefont{F.}~\bibnamefont{Marsili}},
  \bibinfo{author}{\bibfnamefont{V.~B.} \bibnamefont{Verma}},
  \bibinfo{author}{\bibfnamefont{J.~A.} \bibnamefont{Stern}},
  \bibinfo{author}{\bibfnamefont{S.}~\bibnamefont{Harrington}},
  \bibinfo{author}{\bibfnamefont{A.~E.} \bibnamefont{Lita}},
  \bibinfo{author}{\bibfnamefont{T.}~\bibnamefont{Gerrits}},
  \bibinfo{author}{\bibfnamefont{I.}~\bibnamefont{Vayshenker}},
  \bibinfo{author}{\bibfnamefont{B.}~\bibnamefont{Baek}},
  \bibinfo{author}{\bibfnamefont{M.~D.} \bibnamefont{Shaw}},
  \bibinfo{author}{\bibfnamefont{R.~P.} \bibnamefont{Mirin}},
  \bibnamefont{et~al.}, \bibinfo{journal}{Nature Photonics}
  \textbf{\bibinfo{volume}{7}}, \bibinfo{pages}{210} (\bibinfo{year}{2013}).

\bibitem[{\citenamefont{Phan et~al.}(2014)\citenamefont{Phan, Cheng, Bessarab,
  and Krivitsky}}]{frogexperiment}
\bibinfo{author}{\bibfnamefont{N.~M.} \bibnamefont{Phan}},
  \bibinfo{author}{\bibfnamefont{M.~F.} \bibnamefont{Cheng}},
  \bibinfo{author}{\bibfnamefont{D.~A.} \bibnamefont{Bessarab}},
  \bibnamefont{and} \bibinfo{author}{\bibfnamefont{L.~A.}
  \bibnamefont{Krivitsky}}, \bibinfo{journal}{Phys. Rev. Lett.}
  \textbf{\bibinfo{volume}{112}}, \bibinfo{pages}{213601}
  (\bibinfo{year}{2014}).

\bibitem[{\citenamefont{Iskhakov et~al.}(2016)\citenamefont{Iskhakov, Usenko,
  Andersen, Filip, Chekhova, and Leuchs}}]{iskhakov2016heralded}
\bibinfo{author}{\bibfnamefont{T.~S.} \bibnamefont{Iskhakov}},
  \bibinfo{author}{\bibfnamefont{V.~C.} \bibnamefont{Usenko}},
  \bibinfo{author}{\bibfnamefont{U.~L.} \bibnamefont{Andersen}},
  \bibinfo{author}{\bibfnamefont{R.}~\bibnamefont{Filip}},
  \bibinfo{author}{\bibfnamefont{M.~V.} \bibnamefont{Chekhova}},
  \bibnamefont{and} \bibinfo{author}{\bibfnamefont{G.}~\bibnamefont{Leuchs}},
  \bibinfo{journal}{Optics letters} \textbf{\bibinfo{volume}{41}},
  \bibinfo{pages}{2149} (\bibinfo{year}{2016}).

\end{thebibliography}
\end{document}